\documentclass[twocolumn,pra,superscriptaddress,aps,longbibliography]{revtex4-1}
\usepackage{graphicx}
\usepackage[cmex10]{amsmath}
\usepackage[utf8]{inputenc}
\usepackage{float}
\usepackage{placeins}
\usepackage{amssymb}
\usepackage[T1]{fontenc}
\usepackage{hyperref}
\begin{document}

\title{Synchronization properties and reservoir computing capability of hexagonal spintronic oscillator arrays}

\author{Jakub Ch\k{e}ciński}

\affiliation{AGH University of Science and Technology, Department of Electronics, Al. Mickiewicza 30, 30-059 Krak\'{o}w, Poland}

\date{\today}% It is always \today, today,
 %            %  but any date may be explicitly specified

\begin {abstract}

The influence of array geometry on synchronization properties of a 2-D oscillator array is investigated based on a comparison between a rectangular and a hexagonal grid. The Kuramoto model is solved for a nearest-neighbor case with periodic boundary conditions and for a small-scale, realistic coupling case with 1/$\mathrm{r^3}$ decay characteristic of spintronic oscillators. In both cases, it is found that the hexagonal grid choice leads to lower synchronization threshold and higher emission power than its rectangular counterpart, which results from increased connectivity, as well as, in the realistic-coupling case, from decreased contributions of the array edges. Additionally, a more general spin-torque oscillator model including both amplitude and phase as degrees of freedom is employed for reservoir computing simulations, showing that by using hexagonal grid one can increase the short-term memory capacity but not the parity-check capacity of the system.

\end{abstract}

\maketitle

\section{Introduction}

Spintronic oscillators have been drawing a lot of attention as devices that could be utilized for radio-frequency signal generation and transmission as well as for bio-inspired neuromorphic computing \cite{locatelli2014spin,chen2016spin,dieny2019opportunities,choi2014spin,ruiz2017frequency,torrejon2017neuromorphic,romera2018vowel}. Multiple physical implementations, including spin-torque nano-oscillators \cite{kiselev2003microwave,bonetti2009spin,yogendra2015coupled,chen2016spin,torrejon2017neuromorphic,romera2018vowel} ,spin Hall nano-oscillators \cite{liu2012magnetic,demidov2012magnetic,chen2016spin,awad2017long}, oscillators based on antiferromagnets \cite{cheng2016terahertz,khymyn2017antiferromagnetic,khymyn2018ultra} or uncompensated ferrimagnets \cite{lisenkov2019sub} have been proposed, featuring advantages such as the ability to work in room temperature, energy efficiency, small footprint and high frequency ranges being available \cite{locatelli2014spin,chen2016spin,dieny2019opportunities}. The common thread that remains the key to both of the main application areas, namely the DC-to-RF signal conversion and unconventional computing, is making use of oscillator networks and their ability to synchronize by mutually influencing each oscillator's dynamics \cite{locatelli2014spin,chen2016spin,dieny2019opportunities,lebrun2017mutual,tsunegi2018scaling,PhysRevApplied.12.024052}. Therefore, a thorough understanding of the coupling behavior is expected to be crucial for further advancing the spintronic oscillator research.\\

At the same time, synchronization theory in complex networks of oscillators has seen substantial progress within the last several years, creating solid theoretical foundations for understanding most dynamical phenomena occurring in these systems \cite{acebron2005kuramoto,dorfler2014synchronization,rodrigues2016kuramoto}. In this approach, a single oscillator is usually considered to be an abstract object characterized only by its phase, natural frequency and the coupling presence, leading to a simple dynamics equation known as Kuramoto equation \cite{kuramoto2003chemical,acebron2005kuramoto}. Recently, a number of works have shown that Kuramoto equation can be successfully used to describe arrays of spintronic oscillators as well, both in its basic form and in the modified form where the phase degree of freedom is replaced by two degrees of freedom corresponding to oscillation phase and amplitude or power \cite{slavin2009nonlinear,flovik2016describing,PhysRevApplied.12.024052}. Typically, such a description assumes a coupling form that scales with distance as $1/\mathrm{r^3}$, which is characteristic to magnetodipolar coupling that naturally occurs in physical devices such as spin-torque oscillators or spin Hall oscillators \cite{flovik2016describing,PhysRevApplied.12.024052,araujo2015optimizing}.\\

There are, however, important barriers that often make it difficult to directly apply the findings of the oscillation networks theory in the applied spintronics research. Most notably, there exists a vast number of different Kuramoto equation models, from which only a relatively small number can be considered immediately relevant to the spintronic oscillator landscape: many of the commonly made assumptions, such as using periodic boundary conditions, arbitrarily rewired coupling paths in order to increase disorder or an all-to-all coupling scheme, do not hold for a finite-size, magnetically coupled network \cite{watts1998collective,acebron2005kuramoto,dorfler2014synchronization,rodrigues2016kuramoto,hong2005collective,lee2010vortices}. One direction of research that remains relatively unexplored is the influence of grid geometry on synchronization properties. While for an all-to-all coupling scheme the relative physical distances and positions of oscillators can be mostly ignored, in the case of a 2D system with spatially-dependent coupling they can be expected to influence the network dynamics significantly. Despite this, a simple arrangement of oscillators on a cubic Cartesian grid is often assumed in literature by default \cite{sakaguchi1987local,hagerstrom2012experimental,lee2010vortices,flovik2016describing,PhysRevApplied.12.024052}, with the notable exception of the recent works on nanoconstriction-based oscillators, where it was found that the synchronization can be helped by using a tilted rectangular array that assists in spin wave propagation along certain directions. \cite{kendziorczyk2016mutual,zahedinejad2019two}\\

This work investigates synchronization properties and selected neuromorphic computation capabilities for a 2-D oscillator array on a hexagonal grid and compares the obtained results with a rectangular (cubic) grid. Two important cases are examined: firstly, large-scale systems with periodic boundary conditions and nearest-neighbor coupling are considered, in order to highlight some of the differences between the hexagonal arrangement and the results known from the existing synchronization theory literature. Secondly, a more realistic case of a small-scale network with no periodic boundary conditions and $1/\mathrm{r^3}$ coupling decay is analyzed as a direct model of a spintronic oscillators array with magnetic coupling. The synchronization threshold, emission power and short-memory capabilities are all shown to depend on the grid choice significantly, opening new possibilities for oscillator networks research and applications.

\section{Large-scale systems with periodic boundary conditions}

\begin{figure}
\centering
 \includegraphics[width=0.5\textwidth]{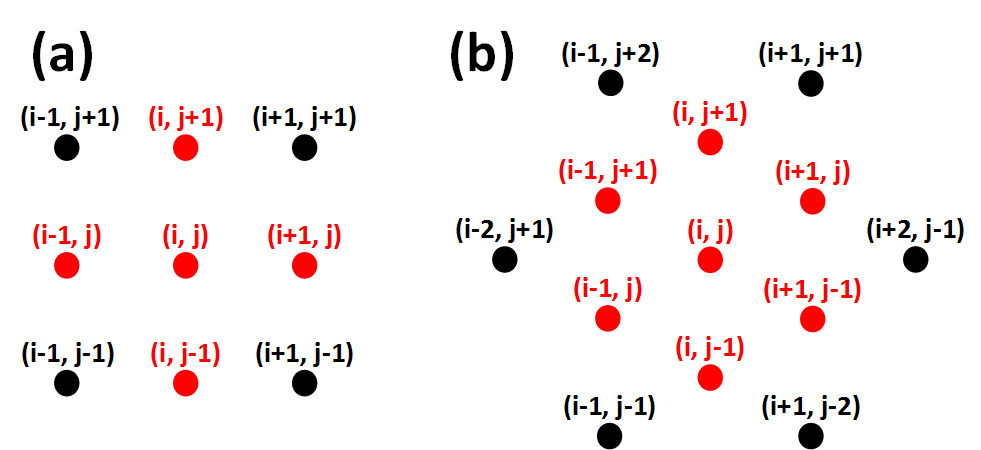}
 \label{fig1}
 \caption{Illustration of rectangular (a) and hexagonal (b) grid coordinates. Red color indicates the central node ${(i,j)}$ and its nearest neighbors.}
\end{figure}

The basic geometrical scheme we examine in this section is pictured in Fig.1, where nearest-neighbors and second-nearest-neighbors of a single grid point $(i,j)$ are shown for the rectangular (a) and the hexagonal (b) grid. Taking into consideration only the nearest-neighbors, the Kuramoto equation for the middle oscillator can be written as \cite{acebron2005kuramoto}:
\begin{align}
&\dot{\phi}_{i,j} = \omega_{i,j} + K\big[sin(\phi_{i-1,j}-\phi_{i,j})+sin(\phi_{i+1,j}-\phi_{i,j})+\nonumber\\
&+sin(\phi_{i,j-1}-\phi_{i,j})+  \nonumber\\ 
&+sin(\phi_{i,j+1}-\phi_{i,j})\big] 
\label{KuramotoRec}
\end{align}
in the rectangular case, where $\phi_{i,j}$ is the phase of the oscillator placed at $(i,j)$ grid point, $\omega_{i,j}$ is its natural frequency and $K$ is the coupling constant. For the hexagonal case, the equation above is modified to include six nearest neighbors, which in hexagonal coordinates can be expressed as:
\begin{align}
&\dot{\phi}_{i,j} = \omega_{i,j} + K\big[sin(\phi_{i-1,j}-\phi_{i,j})+sin(\phi_{i,j-1}-\phi_{i,j})+\nonumber\\
&+sin(\phi_{i-1,j+1}-\phi_{i,j})+sin(\phi_{i,j+1}-\phi_{i,j})+  \nonumber\\ 
&+sin(\phi_{i+1,j-1}-\phi_{i,j})+sin(\phi_{i+1,j}-\phi_{i,j})\big].
\label{Kuramoto}
\end{align}

The analytical expression for the ciritical coupling value $K_{hex}$ necessary to synchronize all the frequencies in the system with a hexagonal grid and periodic boundary conditions will be now investigated. Following the procedure outlined in Ref. \cite{lee2010vortices}, we first consider a linearized version of equation \ref{Kuramoto}:
\begin{equation}
\begin{split}
\dot{\phi}_{i,j} = \omega_i + K(\phi_{i-1,j}+\phi_{i-1,j+1}+\phi_{i,j-1}+\phi_{i,j+1}+ \\ 
+\phi_{i+1,j-1}+\phi_{i+1,j}-6\phi_{i,j}).
\end{split}
\label{LinearKuramoto}
\end{equation}
Solutions of the linearized model can be mapped onto the original Kuramoto model, leading to physically valid solutions of the latter, as long as the phase differences $\Delta\phi$ between oscillator pairs do not exceed a certain threshold value $\Delta_{hex}$ above which the analogy between the two models breaks \cite{lee2010vortices}. In order to determine the exact value of $\Delta_{hex}$, we performed a set of simulations consisting of hexagonal oscillator arrays with periodic boundary conditions. For each linear size of the system $N$, the simulation was repeated 500 times, with oscillator frequencies $\omega_i$ drawn from a Gaussian distribution with unit mean and 0.01 standard deviation, while the uniform coupling constant $K$ was adjusted to ensure that the arrays were on the edge of frequency synchronization. The linear model \ref{LinearKuramoto} was considered a valid approximation of the full Kuramoto model \ref{Kuramoto} for a given set of input frequencies if both of them had a stable solution featuring global frequency synchronization. Our results, summarized in Table \ref{tab:tableDelta}, show that $\Delta_{hex}$ remains in the range of 1.31-1.34 and does not depend on the system size for the analyzed values of $N$. For comparison, we also performed an analogous set of simulations in the rectangular grid case, obtaining $\Delta_{rec}$ typically in the range of 1.22-1.25, which overlaps with the range reported in \cite{lee2010vortices}. \\
\FloatBarrier
\begin{table}
\begin{center}
\begin{tabular}{ |c|c|c| } 
\hline
\multicolumn{1}{|c|}{\bfseries $N$} & \multicolumn{1}{|c|}{\bfseries $\Delta_{rec}$} & \multicolumn{1}{|c|}{\bfseries $\Delta_{hex}$} \\ \hline
 \hspace{10 pt}10\hspace{10 pt} & \hspace{10 pt}1.23 $\pm$ 0.08\hspace{10 pt} & \hspace{10 pt}1.32 $\pm$ 0.10\hspace{10 pt} \\ \hline
 \hspace{10 pt}20\hspace{10 pt} & \hspace{10 pt}1.24 $\pm$ 0.08\hspace{10 pt} & \hspace{10 pt}1.34 $\pm$ 0.08\hspace{10 pt} \\ \hline
 \hspace{10 pt}30\hspace{10 pt} & \hspace{10 pt}1.25 $\pm$ 0.08\hspace{10 pt} & \hspace{10 pt}1.31 $\pm$ 0.08\hspace{10 pt} \\ \hline
 \hspace{10 pt}40\hspace{10 pt} & \hspace{10 pt}1.25 $\pm$ 0.07\hspace{10 pt} & \hspace{10 pt}1.32 $\pm$ 0.08\hspace{10 pt} \\ \hline
 \hspace{10 pt}50\hspace{10 pt} & \hspace{10 pt}1.22 $\pm$ 0.05\hspace{10 pt} & \hspace{10 pt}1.31 $\pm$ 0.09\hspace{10 pt} \\ 
 \hline
\end{tabular}
\caption{Threshold values representing the maximum phase difference in a rectangular ($\Delta_{rec}$) and a hexagonal ($\Delta_{hex}$) grid above which the linear model can no longer be used as an approximation for the full Kuramoto model. The number $N$ represents the linear size of the system.} \label{tab:tableDelta}
\end{center}
\end{table}
%\FloatBarrier
\begin{figure*}
\centering
 \includegraphics[width=1.0\textwidth]{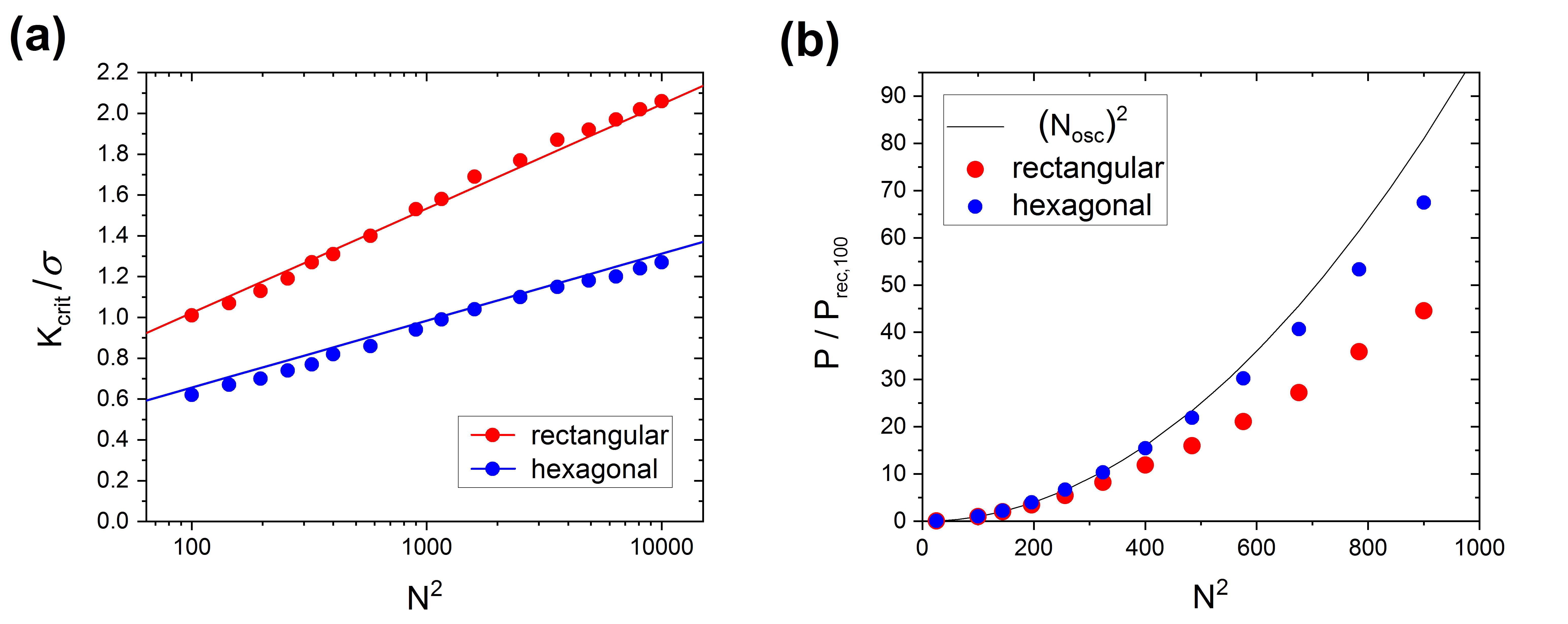}
 \label{fig2}
 \caption{(a) Critical coupling value for synchronization as a function of the total oscillator number $N^2$ for rectangular (red) and hexagonal (blue) grid cases. The dots represent numerical calculations based on Kuramoto equations, the lines represent predictions based on equation \ref{ProbabilitySolved}. (b) Total emission power as a function of the total oscillator number, normalized to the case of rectangular grid with $N^2$ = 100. Red and blue dots refer to numerical results for rectangular and hexagonal systems, respectively, while the black line is drawn as a guide for the eye and represents the ideal $N_{osc}^2$ scaling.}
\end{figure*}
With the link between full and linearized versions of Kuramoto model established, the probability of the global frequency synchronization occurring in the system can be now expressed as the probability of all phase differences  $\Delta\phi$ in the solution of the linearized model being smaller than $\Delta_{hex}$. Using continuum approximation in 2D, it can be shown that $\Delta\phi = \sigma c \cdot \sqrt{\mathrm{log }\ N}/K$, where $\sigma$ is the  standard deviation of frequency and $c$ is a constant that in the case of rectangular grid was calculated analytically as $c_{rec} = \mathrm{1/\sqrt{4\pi}}$ \cite{hong2005collective}. By performing simulations of the linearized model using both types of grid, we have found that $c_{hex} \approx \frac{2}{3} c_{rec}$ for a wide range of $N$ and $K$ values. %, which can be understood intuitively as the effect of the hexagonal grid having $\mathrm{\frac{3}{2}}$ times more connections between oscillators than the rectangular grid for the same system size $N$. 
The synchronization probability, which is the probability of all 3$N^2$ phase differences being simultaneously under the threshold value, can be now expressed as:
\begin{equation}
P(K,N)=\left[\mathrm{erf}\left(\frac{3}{2\sigma}\Delta_{hex}\sqrt{\frac{2\pi}{\mathrm{log}\ N}} K\right)\right]^{3N^2},
\label{Probability}
\end{equation}
where all $\Delta\phi$ are assumed to be mutually independent. In the limit of large $N$, the solution of the above equation for the critical value $K_{crit,hex}$ corresponding to the synchronization threshold will be given by: 
\begin{equation}
\begin{split}
K_{crit,hex} \approx \frac{2\sigma}{3\sqrt{\pi}}\frac{log\ N}{\Delta_{hex}} \approx 0.28 \cdot \sigma log\ N \approx \\
\approx 0.64 \cdot K_{crit,rec}.
\end{split}
\label{ProbabilitySolved}
\end{equation}
We verify the obtained result using a set of 500 numerical simulations of equation \ref{Kuramoto} for both rectangular and hexagonal grids and a wide range of total oscillator numbers $N^2$. For each size, we scanned the coupling constant value $K$ in order to find the critical value $K_{crit}$ corresponding to the synchronization threshold. Because frequency distribution is drawn randomly from the Gaussian distribution each time, different simulations can yield slightly different results for the same set of parameters. Therefore, we define $K_{crit}$ as the value for which more than half of the simulations in the set resulted in all oscillators being fully synchronized in frequency. The results are gathered in Fig.2 (a), where full dots refer to values obtained numerically from the simulations and lines refer to predictions based on equation \ref{ProbabilitySolved}, with red color denoting rectangular grid and blue color denoting hexagonal grid in both cases. Additionally, we present the total oscillation power, defined as $P = \left|\sum_j e^{i\phi_j}\right|^2$ and normalized to the case of rectangular grid with $N^2$ = 100 (Fig.2 (b)). In this set of simulations, the coupling value is kept constant and is equal to $K = 2\sigma$, which means that the system is firmly within the synchronization regime for the chosen range of sizes. Despite this, one can see a deviation from the ideal square dependence on the total oscillator number $N_{osc}$ (please note that since $N$ is the linear size of the array, $(N_{osc})^2=(N^2)^2=N^4$ here). This is caused by increasing phase differences occurring between frequency-synchronized oscillators and is directly affected by the system geometry, as evidenced in Fig.2 (b).
\section{Small-scale systems}
\begin{figure*}
\centering
 \includegraphics[width=1.0\textwidth]{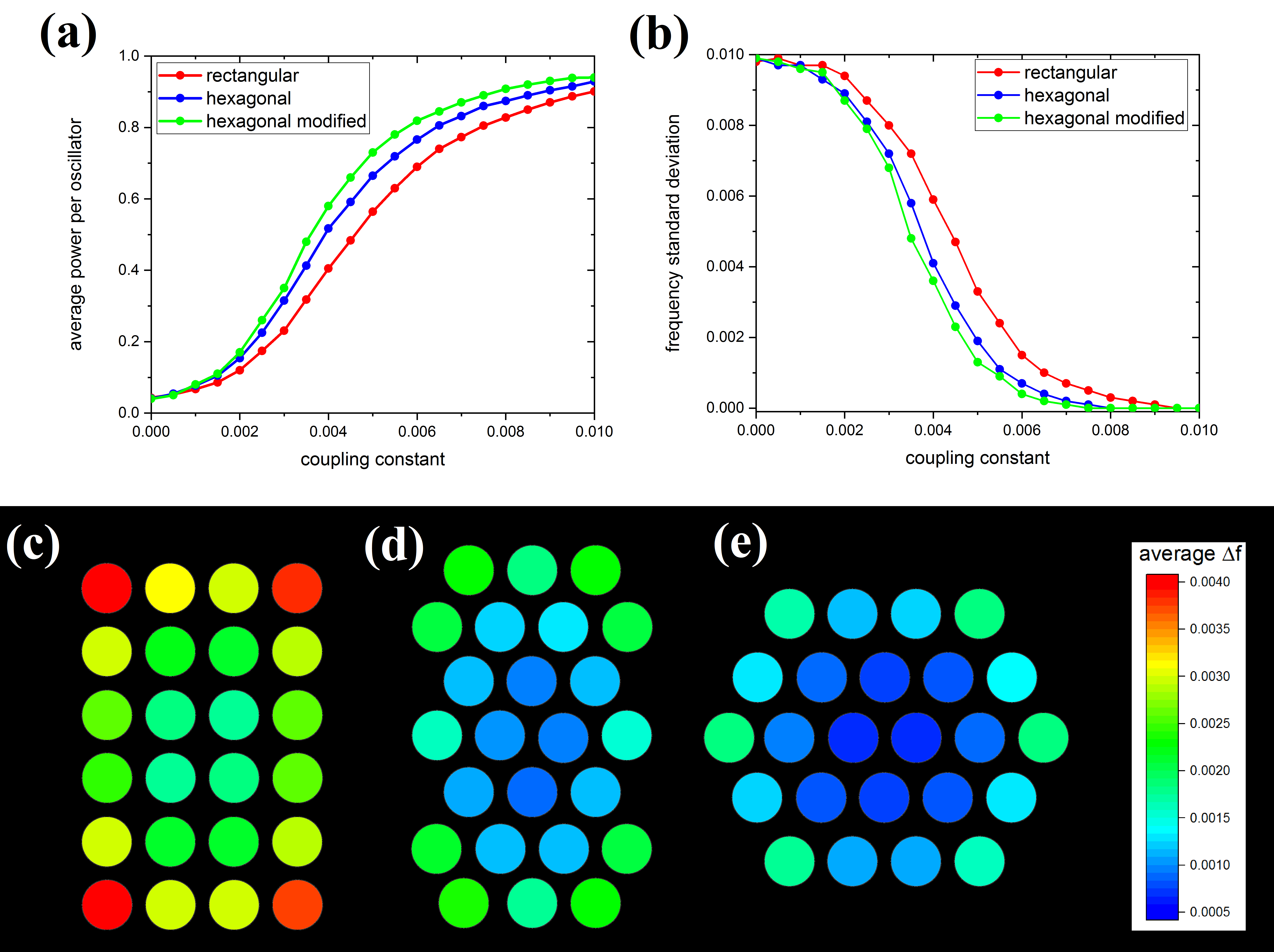}
 \label{fig3}
 \caption{Average power per oscillator (a) and frequency standard deviation (b) as functions of the coupling constant. The lines are drawn as guides for the eye and the points represent three structures considered: rectangular (red), hexagonal (blue) and modified hexagonal (green). The structures are also illustrated below in pictures (c), (d) and (e), respectively, which show the average frequency difference of a single oscillator when compared to the mean frequency of the array.}
\end{figure*}
The results derived in the previous section for large-scale systems with periodic boundary conditions indicate different synchronization behavior in hexagonal arrays of nearest-neighbor Kuramoto oscillators when compared to their rectangular counterparts. However, this alone is not sufficient to make a prediction about the behavior of a small-scale system without any periodic boundary conditions, which is often the most interesting case from the application point of view. We choose to focus here on the specific case of spintronic oscillators that are mutually coupled by a magnetodipolar interaction, which scales with distance as $\mathrm{1/r^3}$. Such a system can be realized as an array of vortex-based oscillators, spin-torque oscillators or spin Hall oscillators and has been shown to feature interesting properties for both synchronization and neuromorphic computing \cite{flovik2016describing, PhysRevApplied.12.024052}. The remainder of this section will describe how a realistic, small-scale system of oscillators with all-to-all magnetodipolar coupling can be affected by the grid geometry choice.\\
\subsection{Frequency deviation and total power}
\begin{figure*}
\centering
 \includegraphics[width=1.0\textwidth]{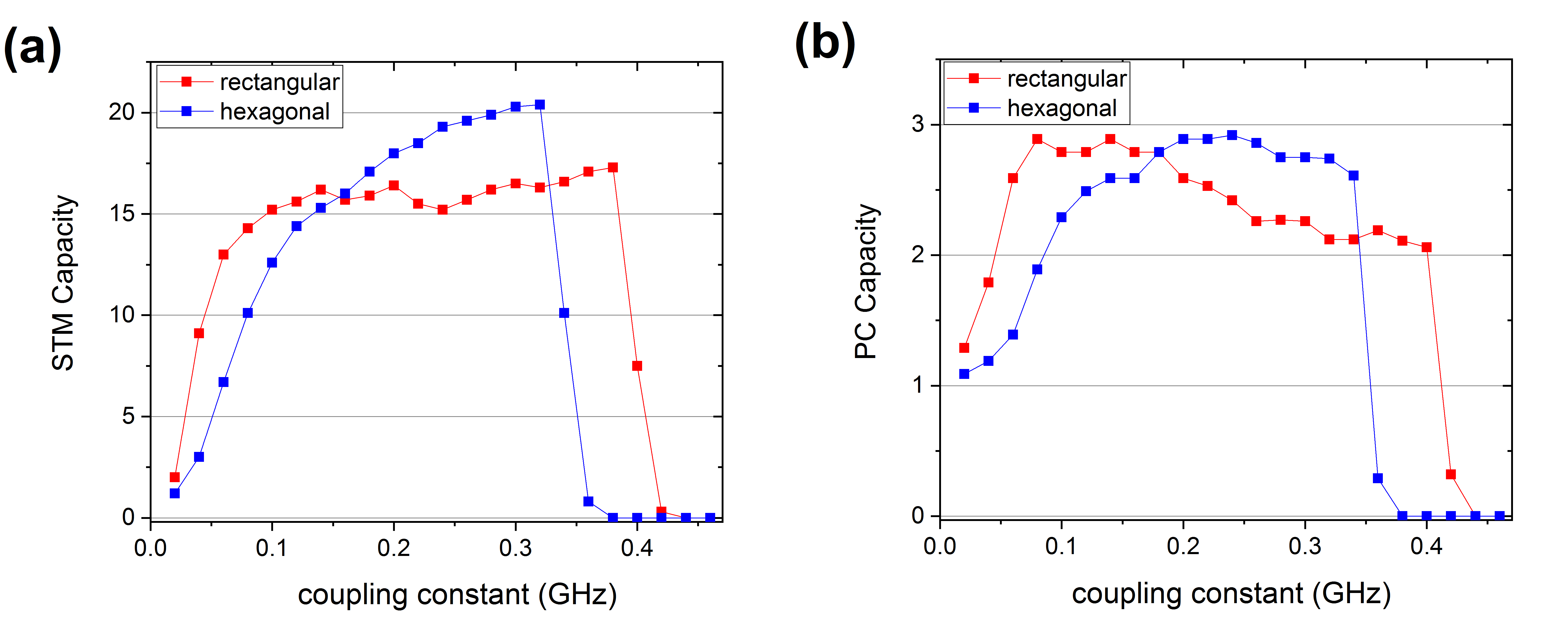}
 \label{fig4}
 \caption{Short-term memory capacity (a) and parity-check capacity (b) of 24-oscillator arrays following rectangular (red color) or hexagonal (blue color) grid geometries. The x axis denotes coupling constant amplitude in GHz. The red and blue lines serve as guides for the eye.}
\end{figure*}
For an example small-scale system with finite boundaries, we choose an array with the total oscillator number $N_{osc}$ = 24, which allows for symmetric placement of oscillators on both rectangular and hexagonal grids. The Kuramoto equations \ref{KuramotoRec} and \ref{Kuramoto} are now replaced by a more general version:
\begin{equation}
\dot{\phi}_{i} = \omega_{i} + \sum_{j\neq i} \frac{K}{r_{ij}^3} sin(\phi_j-\phi_{i}),
\label{Kuramoto2}
\end{equation}
where summing is done over all other oscillators in the array and $r_{ij}$ indicates the geometrical distance between them. One can note that, even though we include all oscillators in the sum to make our analysis more realistic, the steep $1/r^3$ scaling of the coupling constant means that the time evolution of any given $\phi_i$ will be still influenced mostly by its nearest neighbors. Additionally, to make the comparison fair, we will always assume that the distance between any two nearest neighbors is the same in rectangular and in hexagonal grids, so that any differences between them come exclusively from the grid topology. \\

The basic comparison we make is presented in Fig.3 and concerns the average power per oscillator (a) as well as the frequency standard deviation (b) of 24 oscillators arranged in a rectangle that is represented using rectangular (red color) or hexagonal (blue color) coordinates. To understand the origin of the observed difference, we also plot the average frequency difference of a single oscillator when compared to the mean frequency of the array in pictures (c) and (d) for a single coupling constant value $K$ = 0.0045 and $r_{ij}$ = 1 for nearest neighbor pairs. One can see that there are two factors causing the rectangular grid case to underperform: one of them is the influence of the array edges, which are the weakest links from the synchronization point of view, as they have only two nearest neighbors here. Re-arranging the oscillators in a hexagonal manner alleviates this issue, since no oscillator has now less than three nearest neighbors present. However, even in the middle of the array there is a clear difference between both grid cases, which is caused by increased connectivity of hexagonal arrangement that decreases the phase differences between oscillators and thus enhances their ability to synchronize. This can be seen more clearly if we modify the second case to arrange the oscillators in a hexagon shape (Fig.3 (e) and green color in pictures (a) and (b)), which results in further improvement in terms of both power and frequency standard deviation. 
\subsection{Reservoir computing}
To gain further insight into the differences between rectangular and hexagonal grids in the realistic case, we consider the reservoir computing example outlined in references \cite{PhysRevApplied.12.024052,fujii2017harnessing}. The basic system of 24 coupled oscillators was the same as described in the previous section for rectangular and hexagonal grid geometries (Fig.3 (c) and (d)). However, since a typical implementation of reservoir computing in oscillator arrays requires storing information about oscillator amplitudes, it is necessary to modify the original Kuramoto equation from one-variable to two-variables form. We will follow the commonly used approach where each phase $\phi_i$ and amplitude $\sqrt{P_i}$ is expressed as a complex number $c_i = \sqrt{P_i}e^{i\phi_i}$, which results in the following dynamics equation \cite{slavin2008excitation,slavin2009nonlinear,PhysRevApplied.12.024052}:
\begin{equation}
\dot{c_i}=i\omega_i (P_i) c_i-\Gamma_i (P_i) c_i+\sum_{j\neq i}\frac{k}{r_{ij}^3}c_j,
\end{equation}
where each oscillator has its own natural frequency $\omega_i$, $\Gamma$ is the effective damping rate, and $k$ is assumed to be constant so that the coupling strength only depends on the distance between each pair of oscillators \cite{PhysRevApplied.12.024052}. Additionally, we assume the oscillators work in the linear regime where $\omega_i (P_i) = \omega_{0,i} + N(P_i - P_{0,i})$ and $\Gamma_i (P_i) = \Gamma_P(P_i - P_{0,i})/P_{0,i}$, with $N$ and $\Gamma_P$ denoting the nonlinear frequency shift and the dependence of effective damping on oscillator power, respectively.\\

For both systems, we perform the short-term memory (STM) task as well as the parity check (PC) task, which can be used as a proxy for non-linear computations capability \cite{PhysRevApplied.12.024052,fujii2017harnessing}. The parameters we utilized were as following: initial oscillator frequencies $\omega_{0,i}$ drawn randomly from a Gaussian distribution with 3 GHz mean and 50 MHz standard deviation, initial oscillator powers $P_{0,i}$ drawn randomly from a Gaussian distribution with 0.5 mean and 0.01 standard deviation, $N$ = 2$\pi\cdot$ 100 MHz, $\Gamma_P$ = 100 MHz. The coupling constant $k$ was changed between simulations, and the distance between nearest neigbors $r_{ij}$ was always equal to 1. The obtained results, averaged over 100 independent simulations, are gathered in Fig.4, where both the STM capacity (a) and the PC capacity (b) are shown as functions of $k$, with red and blue colors denoting rectangular and hexagonal grids, respectively. It can be seen that the hexagonal case outperforms its rectangular counterpart in term of memory capacity by nearly 20\% in the highest point. Additionally, the coupling value required for maximum STM performance is lower for hexagonal grid, making it easier to achieve in practice. For PC capacity, we do not find any significant difference between the two geometries, indicating that the non-linearity of the considered system stems from inherent non-linear properties of oscillators themselves, and is thus not greatly affected by the grid choice.\\
\section{Summary}
We have demonstrated the influence of grid geometry on selected properties of a 2-D oscillator array with spatially-dependent coupling. By using hexagonal as opposed to cubic lattice for oscillator positions, synchronization threshold was decreased and emission power was increased for both large-scale systems with periodic boundary conditions and for small-scale systems corresponding to a realistic setup of spintronic oscillators such as spin-torque or spin Hall nano-oscillators. Additionally, it has been shown that the hexagonal geometry can also be beneficial in terms of reservoir computing capacity, particularly for tasks relying on short-term memory of the array. The obtained results point towards new directions for theoretical research in complex oscillator networks area and suggest new ways of improving the design of spintronic oscillator arrays. 
\section*{Acknowledgements} Numerical calculations were supported by PL-GRID infrastructure.\\
\bibliography{}

\begin{thebibliography}{0}%
\makeatletter
\providecommand \@ifxundefined [1]{%
 \@ifx{#1\undefined}
}%
\providecommand \@ifnum [1]{%
 \ifnum #1\expandafter \@firstoftwo
 \else \expandafter \@secondoftwo
 \fi
}%
\providecommand \@ifx [1]{%
 \ifx #1\expandafter \@firstoftwo
 \else \expandafter \@secondoftwo
 \fi
}%
\providecommand \natexlab [1]{#1}%
\providecommand \enquote  [1]{``#1''}%
\providecommand \bibnamefont  [1]{#1}%
\providecommand \bibfnamefont [1]{#1}%
\providecommand \citenamefont [1]{#1}%
\providecommand \href@noop [0]{\@secondoftwo}%
\providecommand \href [0]{\begingroup \@sanitize@url \@href}%
\providecommand \@href[1]{\@@startlink{#1}\@@href}%
\providecommand \@@href[1]{\endgroup#1\@@endlink}%
\providecommand \@sanitize@url [0]{\catcode `\\12\catcode `\$12\catcode
  `\&12\catcode `\#12\catcode `\^12\catcode `\_12\catcode `\%12\relax}%
\providecommand \@@startlink[1]{}%
\providecommand \@@endlink[0]{}%
\providecommand \url  [0]{\begingroup\@sanitize@url \@url }%
\providecommand \@url [1]{\endgroup\@href {#1}{\urlprefix }}%
\providecommand \urlprefix  [0]{URL }%
\providecommand \Eprint [0]{\href }%
\providecommand \doibase [0]{http://dx.doi.org/}%
\providecommand \selectlanguage [0]{\@gobble}%
\providecommand \bibinfo  [0]{\@secondoftwo}%
\providecommand \bibfield  [0]{\@secondoftwo}%
\providecommand \translation [1]{[#1]}%
\providecommand \BibitemOpen [0]{}%
\providecommand \bibitemStop [0]{}%
\providecommand \bibitemNoStop [0]{.\EOS\space}%
\providecommand \EOS [0]{\spacefactor3000\relax}%
\providecommand \BibitemShut  [1]{\csname bibitem#1\endcsname}%
\let\auto@bib@innerbib\@empty
%</preamble>
\end{thebibliography}%


\begin{thebibliography}{36}%
\makeatletter
\providecommand \@ifxundefined [1]{%
 \@ifx{#1\undefined}
}%
\providecommand \@ifnum [1]{%
 \ifnum #1\expandafter \@firstoftwo
 \else \expandafter \@secondoftwo
 \fi
}%
\providecommand \@ifx [1]{%
 \ifx #1\expandafter \@firstoftwo
 \else \expandafter \@secondoftwo
 \fi
}%
\providecommand \natexlab [1]{#1}%
\providecommand \enquote  [1]{``#1''}%
\providecommand \bibnamefont  [1]{#1}%
\providecommand \bibfnamefont [1]{#1}%
\providecommand \citenamefont [1]{#1}%
\providecommand \href@noop [0]{\@secondoftwo}%
\providecommand \href [0]{\begingroup \@sanitize@url \@href}%
\providecommand \@href[1]{\@@startlink{#1}\@@href}%
\providecommand \@@href[1]{\endgroup#1\@@endlink}%
\providecommand \@sanitize@url [0]{\catcode `\\12\catcode `\$12\catcode
  `\&12\catcode `\#12\catcode `\^12\catcode `\_12\catcode `\%12\relax}%
\providecommand \@@startlink[1]{}%
\providecommand \@@endlink[0]{}%
\providecommand \url  [0]{\begingroup\@sanitize@url \@url }%
\providecommand \@url [1]{\endgroup\@href {#1}{\urlprefix }}%
\providecommand \urlprefix  [0]{URL }%
\providecommand \Eprint [0]{\href }%
\providecommand \doibase [0]{http://dx.doi.org/}%
\providecommand \selectlanguage [0]{\@gobble}%
\providecommand \bibinfo  [0]{\@secondoftwo}%
\providecommand \bibfield  [0]{\@secondoftwo}%
\providecommand \translation [1]{[#1]}%
\providecommand \BibitemOpen [0]{}%
\providecommand \bibitemStop [0]{}%
\providecommand \bibitemNoStop [0]{.\EOS\space}%
\providecommand \EOS [0]{\spacefactor3000\relax}%
\providecommand \BibitemShut  [1]{\csname bibitem#1\endcsname}%
\let\auto@bib@innerbib\@empty
%</preamble>
\bibitem [{\citenamefont {Locatelli}\ \emph {et~al.}(2014)\citenamefont
  {Locatelli}, \citenamefont {Cros},\ and\ \citenamefont
  {Grollier}}]{locatelli2014spin}%
  \BibitemOpen
  \bibfield  {author} {\bibinfo {author} {\bibfnamefont {N.}~\bibnamefont
  {Locatelli}}, \bibinfo {author} {\bibfnamefont {V.}~\bibnamefont {Cros}}, \
  and\ \bibinfo {author} {\bibfnamefont {J.}~\bibnamefont {Grollier}},\
  }\bibfield  {title} {\enquote {\bibinfo {title} {Spin-torque building
  blocks},}\ }\href@noop {} {\bibfield  {journal} {\bibinfo  {journal} {Nature
  materials}\ }\textbf {\bibinfo {volume} {13}},\ \bibinfo {pages} {11--20}
  (\bibinfo {year} {2014})}\BibitemShut {NoStop}%
\bibitem [{\citenamefont {Chen}\ \emph {et~al.}(2016)\citenamefont {Chen},
  \citenamefont {Dumas}, \citenamefont {Eklund}, \citenamefont {Muduli},
  \citenamefont {Houshang}, \citenamefont {Awad}, \citenamefont
  {D{\"u}rrenfeld}, \citenamefont {Malm}, \citenamefont {Rusu},\ and\
  \citenamefont {{\AA}kerman}}]{chen2016spin}%
  \BibitemOpen
  \bibfield  {author} {\bibinfo {author} {\bibfnamefont {T.}~\bibnamefont
  {Chen}}, \bibinfo {author} {\bibfnamefont {R.~K.}\ \bibnamefont {Dumas}},
  \bibinfo {author} {\bibfnamefont {A.}~\bibnamefont {Eklund}}, \bibinfo
  {author} {\bibfnamefont {P.~K.}\ \bibnamefont {Muduli}}, \bibinfo {author}
  {\bibfnamefont {A.}~\bibnamefont {Houshang}}, \bibinfo {author}
  {\bibfnamefont {A.~A.}\ \bibnamefont {Awad}}, \bibinfo {author}
  {\bibfnamefont {Ph.}\ \bibnamefont {D{\"u}rrenfeld}}, \bibinfo {author}
  {\bibfnamefont {B.~G.}\ \bibnamefont {Malm}}, \bibinfo {author}
  {\bibfnamefont {A.}~\bibnamefont {Rusu}}, \ and\ \bibinfo {author}
  {\bibfnamefont {J.}~\bibnamefont {{\AA}kerman}},\ }\bibfield  {title}
  {\enquote {\bibinfo {title} {Spin-torque and spin-hall nano-oscillators},}\
  }\href@noop {} {\bibfield  {journal} {\bibinfo  {journal} {Proceedings of the
  IEEE}\ }\textbf {\bibinfo {volume} {104}},\ \bibinfo {pages} {1919--1945}
  (\bibinfo {year} {2016})}\BibitemShut {NoStop}%
\bibitem [{\citenamefont {Dieny}\ \emph {et~al.}(2019)\citenamefont {Dieny},
  \citenamefont {Prejbeanu}, \citenamefont {Garello}, \citenamefont
  {Gambardella}, \citenamefont {Freitas}, \citenamefont {Lehndorff},
  \citenamefont {Raberg}, \citenamefont {Ebels}, \citenamefont {Demokritov},
  \citenamefont {{\AA}kerman} \emph {et~al.}}]{dieny2019opportunities}%
  \BibitemOpen
  \bibfield  {author} {\bibinfo {author} {\bibfnamefont {B.}~\bibnamefont
  {Dieny}}, \bibinfo {author} {\bibfnamefont {I.~L.}\ \bibnamefont
  {Prejbeanu}}, \bibinfo {author} {\bibfnamefont {K.}~\bibnamefont {Garello}},
  \bibinfo {author} {\bibfnamefont {P.}~\bibnamefont {Gambardella}}, \bibinfo
  {author} {\bibfnamefont {P.}~\bibnamefont {Freitas}}, \bibinfo {author}
  {\bibfnamefont {R.}~\bibnamefont {Lehndorff}}, \bibinfo {author}
  {\bibfnamefont {W.}~\bibnamefont {Raberg}}, \bibinfo {author} {\bibfnamefont
  {U.}~\bibnamefont {Ebels}}, \bibinfo {author} {\bibfnamefont {S.~O.}\
  \bibnamefont {Demokritov}}, \bibinfo {author} {\bibfnamefont
  {J.}~\bibnamefont {{\AA}kerman}},  \emph {et~al.},\ }\bibfield  {title}
  {\enquote {\bibinfo {title} {Opportunities and challenges for spintronics in
  the microelectronic industry},}\ }\href@noop {} {\bibfield  {journal}
  {\bibinfo  {journal} {arXiv preprint arXiv:1908.10584}\ } (\bibinfo {year}
  {2019})}\BibitemShut {NoStop}%
\bibitem [{\citenamefont {Choi}\ \emph {et~al.}(2014)\citenamefont {Choi},
  \citenamefont {Kang}, \citenamefont {Cho}, \citenamefont {Oh}, \citenamefont
  {Shin}, \citenamefont {Park}, \citenamefont {Jang}, \citenamefont {Min},
  \citenamefont {Kim}, \citenamefont {Park} \emph {et~al.}}]{choi2014spin}%
  \BibitemOpen
  \bibfield  {author} {\bibinfo {author} {\bibfnamefont {H.~S.}\ \bibnamefont
  {Choi}}, \bibinfo {author} {\bibfnamefont {S.~Y.}\ \bibnamefont {Kang}},
  \bibinfo {author} {\bibfnamefont {S.~J.}\ \bibnamefont {Cho}}, \bibinfo
  {author} {\bibfnamefont {I.-Y.}\ \bibnamefont {Oh}}, \bibinfo {author}
  {\bibfnamefont {M.}~\bibnamefont {Shin}}, \bibinfo {author} {\bibfnamefont
  {H.}~\bibnamefont {Park}}, \bibinfo {author} {\bibfnamefont {Ch.}\
  \bibnamefont {Jang}}, \bibinfo {author} {\bibfnamefont {B.-Ch.}\ \bibnamefont
  {Min}}, \bibinfo {author} {\bibfnamefont {S.-Il}\ \bibnamefont {Kim}},
  \bibinfo {author} {\bibfnamefont {S.-Y.}\ \bibnamefont {Park}},  \emph
  {et~al.},\ }\bibfield  {title} {\enquote {\bibinfo {title} {Spin
  nano--oscillator--based wireless communication},}\ }\href@noop {} {\bibfield
  {journal} {\bibinfo  {journal} {Scientific reports}\ }\textbf {\bibinfo
  {volume} {4}},\ \bibinfo {pages} {5486} (\bibinfo {year} {2014})}\BibitemShut
  {NoStop}%
\bibitem [{\citenamefont {Ruiz-Calaforra}\ \emph {et~al.}(2017)\citenamefont
  {Ruiz-Calaforra}, \citenamefont {Purbawati}, \citenamefont {Br{\"a}cher},
  \citenamefont {Hem}, \citenamefont {Murapaka}, \citenamefont {Jim{\'e}nez},
  \citenamefont {Mauri}, \citenamefont {Zeltser}, \citenamefont {Katine},
  \citenamefont {Cyrille} \emph {et~al.}}]{ruiz2017frequency}%
  \BibitemOpen
  \bibfield  {author} {\bibinfo {author} {\bibfnamefont {A.}~\bibnamefont
  {Ruiz-Calaforra}}, \bibinfo {author} {\bibfnamefont {A.}~\bibnamefont
  {Purbawati}}, \bibinfo {author} {\bibfnamefont {T.}~\bibnamefont
  {Br{\"a}cher}}, \bibinfo {author} {\bibfnamefont {J.}~\bibnamefont {Hem}},
  \bibinfo {author} {\bibfnamefont {C.}~\bibnamefont {Murapaka}}, \bibinfo
  {author} {\bibfnamefont {E.}~\bibnamefont {Jim{\'e}nez}}, \bibinfo {author}
  {\bibfnamefont {D.}~\bibnamefont {Mauri}}, \bibinfo {author} {\bibfnamefont
  {A.}~\bibnamefont {Zeltser}}, \bibinfo {author} {\bibfnamefont {J.A.}\
  \bibnamefont {Katine}}, \bibinfo {author} {\bibfnamefont {M.-C.}\
  \bibnamefont {Cyrille}},  \emph {et~al.},\ }\bibfield  {title} {\enquote
  {\bibinfo {title} {Frequency shift keying by current modulation in a
  mtj-based stno with high data rate},}\ }\href@noop {} {\bibfield  {journal}
  {\bibinfo  {journal} {Applied Physics Letters}\ }\textbf {\bibinfo {volume}
  {111}},\ \bibinfo {pages} {082401} (\bibinfo {year} {2017})}\BibitemShut
  {NoStop}%
\bibitem [{\citenamefont {Torrejon}\ \emph {et~al.}(2017)\citenamefont
  {Torrejon}, \citenamefont {Riou}, \citenamefont {Araujo}, \citenamefont
  {Tsunegi}, \citenamefont {Khalsa}, \citenamefont {Querlioz}, \citenamefont
  {Bortolotti}, \citenamefont {Cros}, \citenamefont {Yakushiji}, \citenamefont
  {Fukushima} \emph {et~al.}}]{torrejon2017neuromorphic}%
  \BibitemOpen
  \bibfield  {author} {\bibinfo {author} {\bibfnamefont {J.}~\bibnamefont
  {Torrejon}}, \bibinfo {author} {\bibfnamefont {M.}~\bibnamefont {Riou}},
  \bibinfo {author} {\bibfnamefont {F.~A.}\ \bibnamefont {Araujo}}, \bibinfo
  {author} {\bibfnamefont {S.}~\bibnamefont {Tsunegi}}, \bibinfo {author}
  {\bibfnamefont {G.}~\bibnamefont {Khalsa}}, \bibinfo {author} {\bibfnamefont
  {D.}~\bibnamefont {Querlioz}}, \bibinfo {author} {\bibfnamefont
  {P.}~\bibnamefont {Bortolotti}}, \bibinfo {author} {\bibfnamefont
  {V.}~\bibnamefont {Cros}}, \bibinfo {author} {\bibfnamefont {K.}~\bibnamefont
  {Yakushiji}}, \bibinfo {author} {\bibfnamefont {A.}~\bibnamefont
  {Fukushima}},  \emph {et~al.},\ }\bibfield  {title} {\enquote {\bibinfo
  {title} {Neuromorphic computing with nanoscale spintronic oscillators},}\
  }\href@noop {} {\bibfield  {journal} {\bibinfo  {journal} {Nature}\ }\textbf
  {\bibinfo {volume} {547}},\ \bibinfo {pages} {428} (\bibinfo {year}
  {2017})}\BibitemShut {NoStop}%
\bibitem [{\citenamefont {Romera}\ \emph {et~al.}(2018)\citenamefont {Romera},
  \citenamefont {Talatchian}, \citenamefont {Tsunegi}, \citenamefont {Araujo},
  \citenamefont {Cros}, \citenamefont {Bortolotti}, \citenamefont {Trastoy},
  \citenamefont {Yakushiji}, \citenamefont {Fukushima}, \citenamefont {Kubota}
  \emph {et~al.}}]{romera2018vowel}%
  \BibitemOpen
  \bibfield  {author} {\bibinfo {author} {\bibfnamefont {M.}~\bibnamefont
  {Romera}}, \bibinfo {author} {\bibfnamefont {Ph.}\ \bibnamefont
  {Talatchian}}, \bibinfo {author} {\bibfnamefont {S.}~\bibnamefont {Tsunegi}},
  \bibinfo {author} {\bibfnamefont {F.~A.}\ \bibnamefont {Araujo}}, \bibinfo
  {author} {\bibfnamefont {V.}~\bibnamefont {Cros}}, \bibinfo {author}
  {\bibfnamefont {P.}~\bibnamefont {Bortolotti}}, \bibinfo {author}
  {\bibfnamefont {J.}~\bibnamefont {Trastoy}}, \bibinfo {author} {\bibfnamefont
  {K.}~\bibnamefont {Yakushiji}}, \bibinfo {author} {\bibfnamefont
  {A.}~\bibnamefont {Fukushima}}, \bibinfo {author} {\bibfnamefont
  {H.}~\bibnamefont {Kubota}},  \emph {et~al.},\ }\bibfield  {title} {\enquote
  {\bibinfo {title} {Vowel recognition with four coupled spin-torque
  nano-oscillators},}\ }\href@noop {} {\bibfield  {journal} {\bibinfo
  {journal} {Nature}\ }\textbf {\bibinfo {volume} {563}},\ \bibinfo {pages}
  {230} (\bibinfo {year} {2018})}\BibitemShut {NoStop}%
\bibitem [{\citenamefont {Kiselev}\ \emph {et~al.}(2003)\citenamefont
  {Kiselev}, \citenamefont {Sankey}, \citenamefont {Krivorotov}, \citenamefont
  {Emley}, \citenamefont {Schoelkopf}, \citenamefont {Buhrman},\ and\
  \citenamefont {Ralph}}]{kiselev2003microwave}%
  \BibitemOpen
  \bibfield  {author} {\bibinfo {author} {\bibfnamefont {S.~Il.}\ \bibnamefont
  {Kiselev}}, \bibinfo {author} {\bibfnamefont {J.C.}\ \bibnamefont {Sankey}},
  \bibinfo {author} {\bibfnamefont {I.N.}\ \bibnamefont {Krivorotov}}, \bibinfo
  {author} {\bibfnamefont {N.C.}\ \bibnamefont {Emley}}, \bibinfo {author}
  {\bibfnamefont {R.J.}\ \bibnamefont {Schoelkopf}}, \bibinfo {author}
  {\bibfnamefont {R.A.}\ \bibnamefont {Buhrman}}, \ and\ \bibinfo {author}
  {\bibfnamefont {D.C.}\ \bibnamefont {Ralph}},\ }\bibfield  {title} {\enquote
  {\bibinfo {title} {Microwave oscillations of a nanomagnet driven by a
  spin-polarized current},}\ }\href@noop {} {\bibfield  {journal} {\bibinfo
  {journal} {nature}\ }\textbf {\bibinfo {volume} {425}},\ \bibinfo {pages}
  {380} (\bibinfo {year} {2003})}\BibitemShut {NoStop}%
\bibitem [{\citenamefont {Bonetti}\ \emph {et~al.}(2009)\citenamefont
  {Bonetti}, \citenamefont {Muduli}, \citenamefont {Mancoff},\ and\
  \citenamefont {{\AA}kerman}}]{bonetti2009spin}%
  \BibitemOpen
  \bibfield  {author} {\bibinfo {author} {\bibfnamefont {S.}~\bibnamefont
  {Bonetti}}, \bibinfo {author} {\bibfnamefont {P.}~\bibnamefont {Muduli}},
  \bibinfo {author} {\bibfnamefont {F.}~\bibnamefont {Mancoff}}, \ and\
  \bibinfo {author} {\bibfnamefont {J.}~\bibnamefont {{\AA}kerman}},\
  }\bibfield  {title} {\enquote {\bibinfo {title} {Spin torque oscillator
  frequency versus magnetic field angle: The prospect of operation beyond 65
  ghz},}\ }\href@noop {} {\bibfield  {journal} {\bibinfo  {journal} {Applied
  Physics Letters}\ }\textbf {\bibinfo {volume} {94}},\ \bibinfo {pages}
  {102507} (\bibinfo {year} {2009})}\BibitemShut {NoStop}%
\bibitem [{\citenamefont {Yogendra}\ \emph {et~al.}(2015)\citenamefont
  {Yogendra}, \citenamefont {Fan},\ and\ \citenamefont
  {Roy}}]{yogendra2015coupled}%
  \BibitemOpen
  \bibfield  {author} {\bibinfo {author} {\bibfnamefont {K.}~\bibnamefont
  {Yogendra}}, \bibinfo {author} {\bibfnamefont {D.}~\bibnamefont {Fan}}, \
  and\ \bibinfo {author} {\bibfnamefont {K.}~\bibnamefont {Roy}},\ }\bibfield
  {title} {\enquote {\bibinfo {title} {Coupled spin torque nano oscillators for
  low power neural computation},}\ }\href@noop {} {\bibfield  {journal}
  {\bibinfo  {journal} {IEEE Transactions on Magnetics}\ }\textbf {\bibinfo
  {volume} {51}},\ \bibinfo {pages} {1--9} (\bibinfo {year}
  {2015})}\BibitemShut {NoStop}%
\bibitem [{\citenamefont {Liu}\ \emph {et~al.}(2012)\citenamefont {Liu},
  \citenamefont {Pai}, \citenamefont {Ralph},\ and\ \citenamefont
  {Buhrman}}]{liu2012magnetic}%
  \BibitemOpen
  \bibfield  {author} {\bibinfo {author} {\bibfnamefont {L.}~\bibnamefont
  {Liu}}, \bibinfo {author} {\bibfnamefont {Ch.-F.}\ \bibnamefont {Pai}},
  \bibinfo {author} {\bibfnamefont {D.C.}\ \bibnamefont {Ralph}}, \ and\
  \bibinfo {author} {\bibfnamefont {R.A.}\ \bibnamefont {Buhrman}},\ }\bibfield
   {title} {\enquote {\bibinfo {title} {Magnetic oscillations driven by the
  spin hall effect in 3-terminal magnetic tunnel junction devices},}\
  }\href@noop {} {\bibfield  {journal} {\bibinfo  {journal} {Physical review
  letters}\ }\textbf {\bibinfo {volume} {109}},\ \bibinfo {pages} {186602}
  (\bibinfo {year} {2012})}\BibitemShut {NoStop}%
\bibitem [{\citenamefont {Demidov}\ \emph {et~al.}(2012)\citenamefont
  {Demidov}, \citenamefont {Urazhdin}, \citenamefont {Ulrichs}, \citenamefont
  {Tiberkevich}, \citenamefont {Slavin}, \citenamefont {Baither}, \citenamefont
  {Schmitz},\ and\ \citenamefont {Demokritov}}]{demidov2012magnetic}%
  \BibitemOpen
  \bibfield  {author} {\bibinfo {author} {\bibfnamefont {V.~E.}\ \bibnamefont
  {Demidov}}, \bibinfo {author} {\bibfnamefont {S.}~\bibnamefont {Urazhdin}},
  \bibinfo {author} {\bibfnamefont {H.}~\bibnamefont {Ulrichs}}, \bibinfo
  {author} {\bibfnamefont {V.}~\bibnamefont {Tiberkevich}}, \bibinfo {author}
  {\bibfnamefont {A.}~\bibnamefont {Slavin}}, \bibinfo {author} {\bibfnamefont
  {D.}~\bibnamefont {Baither}}, \bibinfo {author} {\bibfnamefont
  {G.}~\bibnamefont {Schmitz}}, \ and\ \bibinfo {author} {\bibfnamefont
  {S.~O.}\ \bibnamefont {Demokritov}},\ }\bibfield  {title} {\enquote {\bibinfo
  {title} {Magnetic nano-oscillator driven by pure spin current},}\ }\href@noop
  {} {\bibfield  {journal} {\bibinfo  {journal} {Nature materials}\ }\textbf
  {\bibinfo {volume} {11}},\ \bibinfo {pages} {1028} (\bibinfo {year}
  {2012})}\BibitemShut {NoStop}%
\bibitem [{\citenamefont {Awad}\ \emph {et~al.}(2017)\citenamefont {Awad},
  \citenamefont {D{\"u}rrenfeld}, \citenamefont {Houshang}, \citenamefont
  {Dvornik}, \citenamefont {Iacocca}, \citenamefont {Dumas},\ and\
  \citenamefont {{\AA}kerman}}]{awad2017long}%
  \BibitemOpen
  \bibfield  {author} {\bibinfo {author} {\bibfnamefont {A.A.}\ \bibnamefont
  {Awad}}, \bibinfo {author} {\bibfnamefont {Ph.}\ \bibnamefont
  {D{\"u}rrenfeld}}, \bibinfo {author} {\bibfnamefont {A.}~\bibnamefont
  {Houshang}}, \bibinfo {author} {\bibfnamefont {M.}~\bibnamefont {Dvornik}},
  \bibinfo {author} {\bibfnamefont {E.}~\bibnamefont {Iacocca}}, \bibinfo
  {author} {\bibfnamefont {R.K.}\ \bibnamefont {Dumas}}, \ and\ \bibinfo
  {author} {\bibfnamefont {J.}~\bibnamefont {{\AA}kerman}},\ }\bibfield
  {title} {\enquote {\bibinfo {title} {Long-range mutual synchronization of
  spin hall nano-oscillators},}\ }\href@noop {} {\bibfield  {journal} {\bibinfo
   {journal} {Nature Physics}\ }\textbf {\bibinfo {volume} {13}},\ \bibinfo
  {pages} {292} (\bibinfo {year} {2017})}\BibitemShut {NoStop}%
\bibitem [{\citenamefont {Cheng}\ \emph {et~al.}(2016)\citenamefont {Cheng},
  \citenamefont {Xiao},\ and\ \citenamefont {Brataas}}]{cheng2016terahertz}%
  \BibitemOpen
  \bibfield  {author} {\bibinfo {author} {\bibfnamefont {R.}~\bibnamefont
  {Cheng}}, \bibinfo {author} {\bibfnamefont {D.}~\bibnamefont {Xiao}}, \ and\
  \bibinfo {author} {\bibfnamefont {A.}~\bibnamefont {Brataas}},\ }\bibfield
  {title} {\enquote {\bibinfo {title} {Terahertz antiferromagnetic spin hall
  nano-oscillator},}\ }\href@noop {} {\bibfield  {journal} {\bibinfo  {journal}
  {Physical Review Letters}\ }\textbf {\bibinfo {volume} {116}},\ \bibinfo
  {pages} {207603} (\bibinfo {year} {2016})}\BibitemShut {NoStop}%
\bibitem [{\citenamefont {Khymyn}\ \emph {et~al.}(2017)\citenamefont {Khymyn},
  \citenamefont {Lisenkov}, \citenamefont {Tiberkevich}, \citenamefont
  {Ivanov},\ and\ \citenamefont {Slavin}}]{khymyn2017antiferromagnetic}%
  \BibitemOpen
  \bibfield  {author} {\bibinfo {author} {\bibfnamefont {R.}~\bibnamefont
  {Khymyn}}, \bibinfo {author} {\bibfnamefont {I.}~\bibnamefont {Lisenkov}},
  \bibinfo {author} {\bibfnamefont {V.}~\bibnamefont {Tiberkevich}}, \bibinfo
  {author} {\bibfnamefont {B.~A.}\ \bibnamefont {Ivanov}}, \ and\ \bibinfo
  {author} {\bibfnamefont {A.}~\bibnamefont {Slavin}},\ }\bibfield  {title}
  {\enquote {\bibinfo {title} {Antiferromagnetic thz-frequency josephson-like
  oscillator driven by spin current},}\ }\href@noop {} {\bibfield  {journal}
  {\bibinfo  {journal} {Scientific reports}\ }\textbf {\bibinfo {volume} {7}},\
  \bibinfo {pages} {43705} (\bibinfo {year} {2017})}\BibitemShut {NoStop}%
\bibitem [{\citenamefont {Khymyn}\ \emph {et~al.}(2018)\citenamefont {Khymyn},
  \citenamefont {Lisenkov}, \citenamefont {Voorheis}, \citenamefont
  {Sulymenko}, \citenamefont {Prokopenko}, \citenamefont {Tiberkevich},
  \citenamefont {Akerman},\ and\ \citenamefont {Slavin}}]{khymyn2018ultra}%
  \BibitemOpen
  \bibfield  {author} {\bibinfo {author} {\bibfnamefont {R.}~\bibnamefont
  {Khymyn}}, \bibinfo {author} {\bibfnamefont {I.}~\bibnamefont {Lisenkov}},
  \bibinfo {author} {\bibfnamefont {J.}~\bibnamefont {Voorheis}}, \bibinfo
  {author} {\bibfnamefont {O.}~\bibnamefont {Sulymenko}}, \bibinfo {author}
  {\bibfnamefont {O.}~\bibnamefont {Prokopenko}}, \bibinfo {author}
  {\bibfnamefont {V.}~\bibnamefont {Tiberkevich}}, \bibinfo {author}
  {\bibfnamefont {J.}~\bibnamefont {Akerman}}, \ and\ \bibinfo {author}
  {\bibfnamefont {A.}~\bibnamefont {Slavin}},\ }\bibfield  {title} {\enquote
  {\bibinfo {title} {Ultra-fast artificial neuron: generation of
  picosecond-duration spikes in a current-driven antiferromagnetic
  auto-oscillator},}\ }\href@noop {} {\bibfield  {journal} {\bibinfo  {journal}
  {Scientific reports}\ }\textbf {\bibinfo {volume} {8}} (\bibinfo {year}
  {2018})}\BibitemShut {NoStop}%
\bibitem [{\citenamefont {Lisenkov}\ \emph {et~al.}(2019)\citenamefont
  {Lisenkov}, \citenamefont {Khymyn}, \citenamefont {\AA{}kerman},
  \citenamefont {Sun},\ and\ \citenamefont {Ivanov}}]{lisenkov2019sub}%
  \BibitemOpen
  \bibfield  {author} {\bibinfo {author} {\bibfnamefont {Ivan}\ \bibnamefont
  {Lisenkov}}, \bibinfo {author} {\bibfnamefont {Roman}\ \bibnamefont
  {Khymyn}}, \bibinfo {author} {\bibfnamefont {Johan}\ \bibnamefont
  {\AA{}kerman}}, \bibinfo {author} {\bibfnamefont {Nian~X.}\ \bibnamefont
  {Sun}}, \ and\ \bibinfo {author} {\bibfnamefont {Boris~A.}\ \bibnamefont
  {Ivanov}},\ }\bibfield  {title} {\enquote {\bibinfo {title} {Subterahertz
  ferrimagnetic spin-transfer torque oscillator},}\ }\href@noop {} {\bibfield
  {journal} {\bibinfo  {journal} {Phys. Rev. B}\ }\textbf {\bibinfo {volume}
  {100}},\ \bibinfo {pages} {100409} (\bibinfo {year} {2019})}\BibitemShut
  {NoStop}%
\bibitem [{\citenamefont {Lebrun}\ \emph {et~al.}(2017)\citenamefont {Lebrun},
  \citenamefont {Tsunegi}, \citenamefont {Bortolotti}, \citenamefont {Kubota},
  \citenamefont {Jenkins}, \citenamefont {Romera}, \citenamefont {Yakushiji},
  \citenamefont {Fukushima}, \citenamefont {Grollier}, \citenamefont {Yuasa}
  \emph {et~al.}}]{lebrun2017mutual}%
  \BibitemOpen
  \bibfield  {author} {\bibinfo {author} {\bibfnamefont {R.}~\bibnamefont
  {Lebrun}}, \bibinfo {author} {\bibfnamefont {S.}~\bibnamefont {Tsunegi}},
  \bibinfo {author} {\bibfnamefont {P.}~\bibnamefont {Bortolotti}}, \bibinfo
  {author} {\bibfnamefont {H.}~\bibnamefont {Kubota}}, \bibinfo {author}
  {\bibfnamefont {A.S.}\ \bibnamefont {Jenkins}}, \bibinfo {author}
  {\bibfnamefont {M.}~\bibnamefont {Romera}}, \bibinfo {author} {\bibfnamefont
  {K.}~\bibnamefont {Yakushiji}}, \bibinfo {author} {\bibfnamefont
  {A.}~\bibnamefont {Fukushima}}, \bibinfo {author} {\bibfnamefont
  {J.}~\bibnamefont {Grollier}}, \bibinfo {author} {\bibfnamefont
  {S.}~\bibnamefont {Yuasa}},  \emph {et~al.},\ }\bibfield  {title} {\enquote
  {\bibinfo {title} {Mutual synchronization of spin torque nano-oscillators
  through a long-range and tunable electrical coupling scheme},}\ }\href@noop
  {} {\bibfield  {journal} {\bibinfo  {journal} {Nature communications}\
  }\textbf {\bibinfo {volume} {8}},\ \bibinfo {pages} {15825} (\bibinfo {year}
  {2017})}\BibitemShut {NoStop}%
\bibitem [{\citenamefont {Tsunegi}\ \emph {et~al.}(2018)\citenamefont
  {Tsunegi}, \citenamefont {Taniguchi}, \citenamefont {Lebrun}, \citenamefont
  {Yakushiji}, \citenamefont {Cros}, \citenamefont {Grollier}, \citenamefont
  {Fukushima}, \citenamefont {Yuasa},\ and\ \citenamefont
  {Kubota}}]{tsunegi2018scaling}%
  \BibitemOpen
  \bibfield  {author} {\bibinfo {author} {\bibfnamefont {S.}~\bibnamefont
  {Tsunegi}}, \bibinfo {author} {\bibfnamefont {T.}~\bibnamefont {Taniguchi}},
  \bibinfo {author} {\bibfnamefont {R.}~\bibnamefont {Lebrun}}, \bibinfo
  {author} {\bibfnamefont {K.}~\bibnamefont {Yakushiji}}, \bibinfo {author}
  {\bibfnamefont {V.}~\bibnamefont {Cros}}, \bibinfo {author} {\bibfnamefont
  {J.}~\bibnamefont {Grollier}}, \bibinfo {author} {\bibfnamefont
  {A.}~\bibnamefont {Fukushima}}, \bibinfo {author} {\bibfnamefont {Sh.}\
  \bibnamefont {Yuasa}}, \ and\ \bibinfo {author} {\bibfnamefont
  {H.}~\bibnamefont {Kubota}},\ }\bibfield  {title} {\enquote {\bibinfo {title}
  {Scaling up electrically synchronized spin torque oscillator networks},}\
  }\href@noop {} {\bibfield  {journal} {\bibinfo  {journal} {Scientific
  reports}\ }\textbf {\bibinfo {volume} {8}},\ \bibinfo {pages} {13475}
  (\bibinfo {year} {2018})}\BibitemShut {NoStop}%
\bibitem [{\citenamefont {Kanao}\ \emph {et~al.}(2019)\citenamefont {Kanao},
  \citenamefont {Suto}, \citenamefont {Mizushima}, \citenamefont {Goto},
  \citenamefont {Tanamoto},\ and\ \citenamefont
  {Nagasawa}}]{PhysRevApplied.12.024052}%
  \BibitemOpen
  \bibfield  {author} {\bibinfo {author} {\bibfnamefont {T.}~\bibnamefont
  {Kanao}}, \bibinfo {author} {\bibfnamefont {H.}~\bibnamefont {Suto}},
  \bibinfo {author} {\bibfnamefont {K.}~\bibnamefont {Mizushima}}, \bibinfo
  {author} {\bibfnamefont {H.}~\bibnamefont {Goto}}, \bibinfo {author}
  {\bibfnamefont {T.}~\bibnamefont {Tanamoto}}, \ and\ \bibinfo {author}
  {\bibfnamefont {T.}~\bibnamefont {Nagasawa}},\ }\bibfield  {title} {\enquote
  {\bibinfo {title} {Reservoir computing on spin-torque oscillator array},}\
  }\href@noop {} {\bibfield  {journal} {\bibinfo  {journal} {Phys. Rev.
  Applied}\ }\textbf {\bibinfo {volume} {12}},\ \bibinfo {pages} {024052}
  (\bibinfo {year} {2019})}\BibitemShut {NoStop}%
\bibitem [{\citenamefont {Acebr{\'o}n}\ \emph {et~al.}(2005)\citenamefont
  {Acebr{\'o}n}, \citenamefont {Bonilla}, \citenamefont {Vicente},
  \citenamefont {Ritort},\ and\ \citenamefont {Spigler}}]{acebron2005kuramoto}%
  \BibitemOpen
  \bibfield  {author} {\bibinfo {author} {\bibfnamefont {J.~A.}\ \bibnamefont
  {Acebr{\'o}n}}, \bibinfo {author} {\bibfnamefont {L.~L.}\ \bibnamefont
  {Bonilla}}, \bibinfo {author} {\bibfnamefont {C.~J.~P.}\ \bibnamefont
  {Vicente}}, \bibinfo {author} {\bibfnamefont {F.}~\bibnamefont {Ritort}}, \
  and\ \bibinfo {author} {\bibfnamefont {R.}~\bibnamefont {Spigler}},\
  }\bibfield  {title} {\enquote {\bibinfo {title} {The kuramoto model: A simple
  paradigm for synchronization phenomena},}\ }\href@noop {} {\bibfield
  {journal} {\bibinfo  {journal} {Reviews of modern physics}\ }\textbf
  {\bibinfo {volume} {77}},\ \bibinfo {pages} {137} (\bibinfo {year}
  {2005})}\BibitemShut {NoStop}%
\bibitem [{\citenamefont {D{\"o}rfler}\ and\ \citenamefont
  {Bullo}(2014)}]{dorfler2014synchronization}%
  \BibitemOpen
  \bibfield  {author} {\bibinfo {author} {\bibfnamefont {F.}~\bibnamefont
  {D{\"o}rfler}}\ and\ \bibinfo {author} {\bibfnamefont {F.}~\bibnamefont
  {Bullo}},\ }\bibfield  {title} {\enquote {\bibinfo {title} {Synchronization
  in complex networks of phase oscillators: A survey},}\ }\href@noop {}
  {\bibfield  {journal} {\bibinfo  {journal} {Automatica}\ }\textbf {\bibinfo
  {volume} {50}},\ \bibinfo {pages} {1539--1564} (\bibinfo {year}
  {2014})}\BibitemShut {NoStop}%
\bibitem [{\citenamefont {Rodrigues}\ \emph {et~al.}(2016)\citenamefont
  {Rodrigues}, \citenamefont {Peron}, \citenamefont {Ji},\ and\ \citenamefont
  {Kurths}}]{rodrigues2016kuramoto}%
  \BibitemOpen
  \bibfield  {author} {\bibinfo {author} {\bibfnamefont {F.~A.}\ \bibnamefont
  {Rodrigues}}, \bibinfo {author} {\bibfnamefont {Th.~K.D.M.}\ \bibnamefont
  {Peron}}, \bibinfo {author} {\bibfnamefont {P.}~\bibnamefont {Ji}}, \ and\
  \bibinfo {author} {\bibfnamefont {J.}~\bibnamefont {Kurths}},\ }\bibfield
  {title} {\enquote {\bibinfo {title} {The kuramoto model in complex
  networks},}\ }\href@noop {} {\bibfield  {journal} {\bibinfo  {journal}
  {Physics Reports}\ }\textbf {\bibinfo {volume} {610}},\ \bibinfo {pages}
  {1--98} (\bibinfo {year} {2016})}\BibitemShut {NoStop}%
\bibitem [{\citenamefont {Kuramoto}(2003)}]{kuramoto2003chemical}%
  \BibitemOpen
  \bibfield  {author} {\bibinfo {author} {\bibfnamefont {Y.}~\bibnamefont
  {Kuramoto}},\ }\href@noop {} {\emph {\bibinfo {title} {Chemical oscillations,
  waves, and turbulence}}}\ (\bibinfo  {publisher} {Courier Corporation},\
  \bibinfo {year} {2003})\BibitemShut {NoStop}%
\bibitem [{\citenamefont {Slavin}\ and\ \citenamefont
  {Tiberkevich}(2009)}]{slavin2009nonlinear}%
  \BibitemOpen
  \bibfield  {author} {\bibinfo {author} {\bibfnamefont {A.}~\bibnamefont
  {Slavin}}\ and\ \bibinfo {author} {\bibfnamefont {V.}~\bibnamefont
  {Tiberkevich}},\ }\bibfield  {title} {\enquote {\bibinfo {title} {Nonlinear
  auto-oscillator theory of microwave generation by spin-polarized current},}\
  }\href@noop {} {\bibfield  {journal} {\bibinfo  {journal} {IEEE Transactions
  on Magnetics}\ }\textbf {\bibinfo {volume} {45}},\ \bibinfo {pages}
  {1875--1918} (\bibinfo {year} {2009})}\BibitemShut {NoStop}%
\bibitem [{\citenamefont {Flovik}\ \emph {et~al.}(2016)\citenamefont {Flovik},
  \citenamefont {Macia},\ and\ \citenamefont
  {Wahlstr{\"o}m}}]{flovik2016describing}%
  \BibitemOpen
  \bibfield  {author} {\bibinfo {author} {\bibfnamefont {V.}~\bibnamefont
  {Flovik}}, \bibinfo {author} {\bibfnamefont {F.}~\bibnamefont {Macia}}, \
  and\ \bibinfo {author} {\bibfnamefont {E.}~\bibnamefont {Wahlstr{\"o}m}},\
  }\bibfield  {title} {\enquote {\bibinfo {title} {Describing synchronization
  and topological excitations in arrays of magnetic spin torque oscillators
  through the kuramoto model},}\ }\href@noop {} {\bibfield  {journal} {\bibinfo
   {journal} {Scientific reports}\ }\textbf {\bibinfo {volume} {6}},\ \bibinfo
  {pages} {32528} (\bibinfo {year} {2016})}\BibitemShut {NoStop}%
\bibitem [{\citenamefont {Araujo}\ \emph {et~al.}(2015)\citenamefont {Araujo},
  \citenamefont {Belanovsky}, \citenamefont {Skirdkov}, \citenamefont
  {Zvezdin}, \citenamefont {Zvezdin}, \citenamefont {Locatelli}, \citenamefont
  {Lebrun}, \citenamefont {Grollier}, \citenamefont {Cros}, \citenamefont
  {De~Loubens} \emph {et~al.}}]{araujo2015optimizing}%
  \BibitemOpen
  \bibfield  {author} {\bibinfo {author} {\bibfnamefont {F.~A.}\ \bibnamefont
  {Araujo}}, \bibinfo {author} {\bibfnamefont {A.D.}\ \bibnamefont
  {Belanovsky}}, \bibinfo {author} {\bibfnamefont {P.N.}\ \bibnamefont
  {Skirdkov}}, \bibinfo {author} {\bibfnamefont {K.A.}\ \bibnamefont
  {Zvezdin}}, \bibinfo {author} {\bibfnamefont {A.K.}\ \bibnamefont {Zvezdin}},
  \bibinfo {author} {\bibfnamefont {N.}~\bibnamefont {Locatelli}}, \bibinfo
  {author} {\bibfnamefont {R.}~\bibnamefont {Lebrun}}, \bibinfo {author}
  {\bibfnamefont {J.}~\bibnamefont {Grollier}}, \bibinfo {author}
  {\bibfnamefont {V.}~\bibnamefont {Cros}}, \bibinfo {author} {\bibfnamefont
  {G.}~\bibnamefont {De~Loubens}},  \emph {et~al.},\ }\bibfield  {title}
  {\enquote {\bibinfo {title} {Optimizing magnetodipolar interactions for
  synchronizing vortex based spin-torque nano-oscillators},}\ }\href@noop {}
  {\bibfield  {journal} {\bibinfo  {journal} {Physical Review B}\ }\textbf
  {\bibinfo {volume} {92}},\ \bibinfo {pages} {045419} (\bibinfo {year}
  {2015})}\BibitemShut {NoStop}%
\bibitem [{\citenamefont {Watts}\ and\ \citenamefont
  {Strogatz}(1998)}]{watts1998collective}%
  \BibitemOpen
  \bibfield  {author} {\bibinfo {author} {\bibfnamefont {D.~J.}\ \bibnamefont
  {Watts}}\ and\ \bibinfo {author} {\bibfnamefont {S.~H.}\ \bibnamefont
  {Strogatz}},\ }\bibfield  {title} {\enquote {\bibinfo {title} {Collective
  dynamics of ‘small-world’networks},}\ }\href@noop {} {\bibfield
  {journal} {\bibinfo  {journal} {nature}\ }\textbf {\bibinfo {volume} {393}},\
  \bibinfo {pages} {440} (\bibinfo {year} {1998})}\BibitemShut {NoStop}%
\bibitem [{\citenamefont {Hong}\ \emph {et~al.}(2005)\citenamefont {Hong},
  \citenamefont {Park},\ and\ \citenamefont {Choi}}]{hong2005collective}%
  \BibitemOpen
  \bibfield  {author} {\bibinfo {author} {\bibfnamefont {H.}~\bibnamefont
  {Hong}}, \bibinfo {author} {\bibfnamefont {H.}~\bibnamefont {Park}}, \ and\
  \bibinfo {author} {\bibfnamefont {M.Y.}\ \bibnamefont {Choi}},\ }\bibfield
  {title} {\enquote {\bibinfo {title} {Collective synchronization in spatially
  extended systems of coupled oscillators with random frequencies},}\
  }\href@noop {} {\bibfield  {journal} {\bibinfo  {journal} {Physical Review
  E}\ }\textbf {\bibinfo {volume} {72}},\ \bibinfo {pages} {036217} (\bibinfo
  {year} {2005})}\BibitemShut {NoStop}%
\bibitem [{\citenamefont {Lee}\ \emph {et~al.}(2010)\citenamefont {Lee},
  \citenamefont {Tam}, \citenamefont {Refael}, \citenamefont {Rogers},\ and\
  \citenamefont {Cross}}]{lee2010vortices}%
  \BibitemOpen
  \bibfield  {author} {\bibinfo {author} {\bibfnamefont {T.~E.}\ \bibnamefont
  {Lee}}, \bibinfo {author} {\bibfnamefont {H.}~\bibnamefont {Tam}}, \bibinfo
  {author} {\bibfnamefont {G.}~\bibnamefont {Refael}}, \bibinfo {author}
  {\bibfnamefont {J.~L.}\ \bibnamefont {Rogers}}, \ and\ \bibinfo {author}
  {\bibfnamefont {M.~C.}\ \bibnamefont {Cross}},\ }\bibfield  {title} {\enquote
  {\bibinfo {title} {Vortices and the entrainment transition in the
  two-dimensional kuramoto model},}\ }\href@noop {} {\bibfield  {journal}
  {\bibinfo  {journal} {Physical Review E}\ }\textbf {\bibinfo {volume} {82}},\
  \bibinfo {pages} {036202} (\bibinfo {year} {2010})}\BibitemShut {NoStop}%
\bibitem [{\citenamefont {Sakaguchi}\ \emph {et~al.}(1987)\citenamefont
  {Sakaguchi}, \citenamefont {Shinomoto},\ and\ \citenamefont
  {Kuramoto}}]{sakaguchi1987local}%
  \BibitemOpen
  \bibfield  {author} {\bibinfo {author} {\bibfnamefont {H.}~\bibnamefont
  {Sakaguchi}}, \bibinfo {author} {\bibfnamefont {Sh.}\ \bibnamefont
  {Shinomoto}}, \ and\ \bibinfo {author} {\bibfnamefont {Y.}~\bibnamefont
  {Kuramoto}},\ }\bibfield  {title} {\enquote {\bibinfo {title} {Local and
  grobal self-entrainments in oscillator lattices},}\ }\href@noop {} {\bibfield
   {journal} {\bibinfo  {journal} {Progress of Theoretical Physics}\ }\textbf
  {\bibinfo {volume} {77}},\ \bibinfo {pages} {1005--1010} (\bibinfo {year}
  {1987})}\BibitemShut {NoStop}%
\bibitem [{\citenamefont {Hagerstrom}\ \emph {et~al.}(2012)\citenamefont
  {Hagerstrom}, \citenamefont {Murphy}, \citenamefont {Roy}, \citenamefont
  {H{\"o}vel}, \citenamefont {Omelchenko},\ and\ \citenamefont
  {Sch{\"o}ll}}]{hagerstrom2012experimental}%
  \BibitemOpen
  \bibfield  {author} {\bibinfo {author} {\bibfnamefont {A.~M.}\ \bibnamefont
  {Hagerstrom}}, \bibinfo {author} {\bibfnamefont {Th.~E.}\ \bibnamefont
  {Murphy}}, \bibinfo {author} {\bibfnamefont {R.}~\bibnamefont {Roy}},
  \bibinfo {author} {\bibfnamefont {Ph.}\ \bibnamefont {H{\"o}vel}}, \bibinfo
  {author} {\bibfnamefont {I.}~\bibnamefont {Omelchenko}}, \ and\ \bibinfo
  {author} {\bibfnamefont {E.}~\bibnamefont {Sch{\"o}ll}},\ }\bibfield  {title}
  {\enquote {\bibinfo {title} {Experimental observation of chimeras in
  coupled-map lattices},}\ }\href@noop {} {\bibfield  {journal} {\bibinfo
  {journal} {Nature Physics}\ }\textbf {\bibinfo {volume} {8}},\ \bibinfo
  {pages} {658} (\bibinfo {year} {2012})}\BibitemShut {NoStop}%
\bibitem [{\citenamefont {Kendziorczyk}\ and\ \citenamefont
  {Kuhn}(2016)}]{kendziorczyk2016mutual}%
  \BibitemOpen
  \bibfield  {author} {\bibinfo {author} {\bibfnamefont {T.}~\bibnamefont
  {Kendziorczyk}}\ and\ \bibinfo {author} {\bibfnamefont {T.}~\bibnamefont
  {Kuhn}},\ }\bibfield  {title} {\enquote {\bibinfo {title} {Mutual
  synchronization of nanoconstriction-based spin hall nano-oscillators through
  evanescent and propagating spin waves},}\ }\href@noop {} {\bibfield
  {journal} {\bibinfo  {journal} {Physical Review B}\ }\textbf {\bibinfo
  {volume} {93}},\ \bibinfo {pages} {134413} (\bibinfo {year}
  {2016})}\BibitemShut {NoStop}%
\bibitem [{\citenamefont {Zahedinejad}\ \emph {et~al.}(2019)\citenamefont
  {Zahedinejad}, \citenamefont {Awad}, \citenamefont {Muralidhar},
  \citenamefont {Khymyn}, \citenamefont {Fulara}, \citenamefont {Mazraati},
  \citenamefont {Dvornik},\ and\ \citenamefont
  {{\AA}kerman}}]{zahedinejad2019two}%
  \BibitemOpen
  \bibfield  {author} {\bibinfo {author} {\bibfnamefont {Mohammad}\
  \bibnamefont {Zahedinejad}}, \bibinfo {author} {\bibfnamefont {Ahmad~A}\
  \bibnamefont {Awad}}, \bibinfo {author} {\bibfnamefont {Shreyas}\
  \bibnamefont {Muralidhar}}, \bibinfo {author} {\bibfnamefont {Roman}\
  \bibnamefont {Khymyn}}, \bibinfo {author} {\bibfnamefont {Himanshu}\
  \bibnamefont {Fulara}}, \bibinfo {author} {\bibfnamefont {Hamid}\
  \bibnamefont {Mazraati}}, \bibinfo {author} {\bibfnamefont {Mykola}\
  \bibnamefont {Dvornik}}, \ and\ \bibinfo {author} {\bibfnamefont {Johan}\
  \bibnamefont {{\AA}kerman}},\ }\bibfield  {title} {\enquote {\bibinfo {title}
  {Two-dimensional mutually synchronized spin hall nano-oscillator arrays for
  neuromorphic computing},}\ }\href@noop {} {\bibfield  {journal} {\bibinfo
  {journal} {Nature Nanotechnology}\ ,\ \bibinfo {pages} {1--6}} (\bibinfo
  {year} {2019})}\BibitemShut {NoStop}%
\bibitem [{\citenamefont {Fujii}\ and\ \citenamefont
  {Nakajima}(2017)}]{fujii2017harnessing}%
  \BibitemOpen
  \bibfield  {author} {\bibinfo {author} {\bibfnamefont {K.}~\bibnamefont
  {Fujii}}\ and\ \bibinfo {author} {\bibfnamefont {K.}~\bibnamefont
  {Nakajima}},\ }\bibfield  {title} {\enquote {\bibinfo {title} {Harnessing
  disordered-ensemble quantum dynamics for machine learning},}\ }\href@noop {}
  {\bibfield  {journal} {\bibinfo  {journal} {Physical Review Applied}\
  }\textbf {\bibinfo {volume} {8}},\ \bibinfo {pages} {024030} (\bibinfo {year}
  {2017})}\BibitemShut {NoStop}%
\bibitem [{\citenamefont {Slavin}\ and\ \citenamefont
  {Tiberkevich}(2008)}]{slavin2008excitation}%
  \BibitemOpen
  \bibfield  {author} {\bibinfo {author} {\bibfnamefont {A.}~\bibnamefont
  {Slavin}}\ and\ \bibinfo {author} {\bibfnamefont {V.}~\bibnamefont
  {Tiberkevich}},\ }\bibfield  {title} {\enquote {\bibinfo {title} {Excitation
  of spin waves by spin-polarized current in magnetic nano-structures},}\
  }\href@noop {} {\bibfield  {journal} {\bibinfo  {journal} {IEEE Transactions
  on Magnetics}\ }\textbf {\bibinfo {volume} {44}},\ \bibinfo {pages}
  {1916--1927} (\bibinfo {year} {2008})}\BibitemShut {NoStop}%
\end{thebibliography}%

\end{document}